\documentclass{article}
\usepackage{amsfonts}
\usepackage{amsmath}
\usepackage{epsfig}
\usepackage{comment}

\begin{document}

\title{Optics in  nonuniform  media \\ and Lagrange geometry}
\author{M. Neagu, N. G. Krylova, E. M. Ovsiyuk, V. M. Red'kov}
\date{}
\maketitle

\begin{abstract}
In this paper the equations of motion  associated with a
Lagrangian inspired by relativistic optics in nonuniform moving
media are considered. The model describes optical effects in the
nonuniform dispersionless moving medium. When using the optical
metric restricted to the Minkowski manifold, we have established
the Euler-Lagrange equations  for geodesics. We have specified the
general model to the special case when the refractive index
increases along the direction $Z$.  The exact analytical solutions
of the corresponding Euler-Lagrange equations have been
constructed.  Analysis of the solutions shows that the light beams
are bending to the axes $Z$ along which the refractive index
increases.
\end{abstract}


\section{Introduction}

In geometrical optics \cite{Landau-Lif-1}, a special role is played by the
Beil metric (see 
\cite{An-Shim}-\cite{Beil}, \cite{Gordon}, \cite{Leonard}-\cite{Szasz})
\begin{equation}
g_{\alpha \beta }(x,y)=\varphi _{\alpha \beta }(x)+\gamma ^{2}(x)y_{\alpha
}y_{\beta },  \label{Synge-Beil}
\end{equation}%
where $\gamma (x)\geq 0$ is a  smooth function on the space-time $%
M^{4}$, and $\varphi _{\alpha \beta }(x)$ is a pseudo-Riemannian metric on $%
M^{4}$. One assumes that the manifold $M^{4}$ 
is endowed with local coordinates $%
\left( x^{\alpha }\right) _{\alpha =\overline{1,4}}=\left(
x^{1}=t,x^{2},x^{3},x^{4}\right) $, and $\left( y^{\alpha }\right)$ 
is the Liouville vector field on the total space of the
tangent bundle $TM^{4}$; 
the following rule holds $%
y_{\alpha }=\varphi _{\alpha \mu }y^{\mu }$. Since the components of $%
\varphi _{\alpha \beta }(x)$ are dimensionless, the same are
$\gamma y_{\alpha }$; so we have dimensionless combinations
$[\varphi _{\alpha \beta }(x)]=1,\;[\gamma y_{\alpha }]=1.$

In this context, let us restrict our 
study to the
Minkowski manifold $\mathcal{M}^{4}=\left( \mathbb{R}^{4},\eta _{ij}\right) $
which has the local coordinates $(x):=\left( x^{i}\right) _{i=\overline{1,4}%
}.$ The dimension of the corresponding tangent bundle $T%
\mathbb{R}^{4}$ is equal to eight, and its local coordinates are\footnote{%
In this paper, the Latin letters $i,$ $j,$ $k,$ ... run from $1$
to $4$. The Einstein convention of summation is adopted. In
order to eliminate the confusion between indices and powers, for the space-time coordinates we will use  the notations: $x^{i}:=x^{(i)},$ $\forall $ $i=%
\overline{1,4}.$}%
\begin{equation*}
(x,y):=(x^{i},y^{i})_{i=\overline{1,4}}=(\underset{\text{space-time
coordinates}}{\underbrace{x^{1},x^{2},x^{3},x^{4}}},\text{ }\underset{\text{%
tangent vector}}{\underbrace{y^{1},y^{2},y^{3},y^{4}}}\quad ).
\end{equation*}%
Emerging from formula (\ref{Synge-Beil}), we introduce the
following metric on $T\mathbb{R}^{4}$, which is inspired by the
optics framework developed in the papers
\cite{Gordon},\cite{Mandelstam},\cite{Mir-Kaw},\cite{Synge}-\cite{Tamm2}
for the nonuniform moving medium:
\begin{equation}
\mathfrak{g}_{ij}(x,y)=\eta _{ij}+\gamma ^{2}(x)y_{i}y_{j},
\label{Minkovski-optics}
\end{equation}%
where $\eta =\left( \eta _{ij}\right) =$ diag $(-1,1,1,1)$ is the Minkowski
metric, and $y_{i}=\eta _{ir}y^{r}$. Commonly, the following parametrization for $\gamma ^{2}(x)$ is used%
\begin{equation}
\gamma^{2}(x)=\frac{1}{c^{2}}\left( 1-\frac{1}{n^{2}(x)}\right),
\label{gamman}
\end{equation}%
where $n=n(x)$ is interpreted as the local refractive index of the nonuniform medium (see \cite%
{Balan-Synge}, \cite{Landau-Lif-1}, \cite{Mir-Kaw}-\cite{NOR}).

From the physical point of view, geometrical optics in moving
media is an interesting object because the effects of velocity
vector field are similar to the action of gravitational  or
magnetic fields on charged matter waves. In paper \cite{Leonard},
the Lagrangian and the metric related to the light in moving
dispersionless media have been established and the
gravitation-like effects for the light deflection at a vortex has
been studied (so called an optical black hole). It should be noted
that in such models \cite{Leonard} the second term in the Gordon's optical metric \cite{Gordon}
\begin{equation}
g_{ij}=\eta _{ij}+(1-n^{-2})u_{i}u_{j}  \label{SyngeMetric}
\end{equation}%
describes medium velocity effects; here the four-velocity is defined by the formula%
\begin{equation*}
u_{i}(x)=\alpha \left( 1,-\frac{\mathbf{u}}{c}\right), \qquad
\alpha =\left(1-\frac{u^2}{c^2}\right)^{-1/2}.
\end{equation*}%

In the sequel, 
we will examine the special case of an anisotropic
dynamical model, 
which is governed by the Lagrangian 
\cite{Neagu+Oana Optics}:
\begin{equation}
L(x,y)  =  \dfrac{1}{2}\mathfrak{g}_{ij}(x,y)y^{i}y^{j}  =
\dfrac{1}{2}(\eta _{ij}+\gamma ^{2}y_{i}y_{j})y^{i}y^{j}
 =  \dfrac{1}{2}\eta _{ij}y^{i}y^{j}+{\dfrac{\gamma ^{2}}{2}}||y||^{4},%
\label{Lagragian-ARO}
\end{equation}%
where the notation is used
$$||y||^{2}=-(y^{1})^{2}+(y^{2})^{2}+(y^{3})^{2}+(y^{4})^{2}=\eta
_{ij}y^{i}y^{j}.$$

Assuming that the refractive index $n(x)$ is invariant with
respect to Lorentz transformations, we conclude that the
Lagrangian (\ref{Lagragian-ARO}) is also invariant. A similar
3-dimensional Lagrangian 
was studied in \cite{NOR}, the corresponding non-relativistic
 Lagrangian being invariant with respect to the orthogonal
group $O(3)$. 

The Lagrangian (\ref{Lagragian-ARO}) produces the fundamental
metric
\begin{equation*}
g_{ij}(x,y)=\frac{1}{2}\frac{\partial ^{2}L}{\partial y^{i}\partial y^{j}}%
=\sigma (x,y)\eta _{ij}+2\gamma ^{2}(x)y_{i}y_{j},
\end{equation*}%
where $\sigma (x,y)=(1/2)+\gamma ^{2}(x)||y||^{2}.$
With notation $\tau (x,y)=(1/2)+3\gamma ^{2}(x)||y||^{2}$,  for
the inverse matrix $[g^{-1}]=(g^{jk})_{j,k=\overline{1,4}}$ we
find
\begin{equation*}
g^{jk}(x,y)=\frac{1}{\sigma (x,y)}\eta ^{jk}-\frac{2\gamma ^{2}(x)}{\sigma
(x,y)\cdot \tau (x,y)}y^{j}y^{k},
\end{equation*}%
where we assume that  $\sigma \cdot \tau \neq 0$. The
Euler-Lagrange equations associated with the Lagrangian
(\ref{Lagragian-ARO}) can be written in the form  (see
\cite{Balan-Synge})
\begin{equation}
\frac{d^{2}x^{i}}{dt^{2}}+2G^{i}\left( x\left( t\right) ,y\left( t\right)
\right) =0,  \label{Euler-Lagrange=0}
\end{equation}%
where $G^{i}$ is defined by the formula
$$
G^{i}(x,y)\overset{def}{=}\frac{g^{ik}}{4}\left( \frac{\partial ^{2}L}{%
\partial y^{k}\partial x^{s}}y^{s}-\frac{\partial L}{\partial x^{k}}\right)
$$
$$=\frac{\gamma }{\sigma }\left\vert \left\vert y\right\vert \right\vert
^{2}y^{i}\left( \gamma _{s}y^{s}\right) -\frac{3}{2}\frac{\gamma ^{3}}{%
\sigma \tau }y^{i}\left\vert \left\vert y\right\vert \right\vert ^{4}\left(
\gamma _{s}y^{s}\right) -\frac{\gamma }{4\sigma }\left\vert \left\vert
y\right\vert \right\vert ^{4}\gamma _{i}\eta _{ii},
$$
here $\gamma _{s}=\partial \gamma /\partial x^{s}$.
The geometrical quantity $G^{i}(x,y)$ has the meaning
of the semispray on the tangent space $T\mathbb{R}^{4}$ (see \cite{Mir-An}, \cite{Bal-Nea-Wiley}).

\section{The Euler-Lagrange equations}

The Euler-Lagrange equations (\ref{Euler-Lagrange=0}) for the variables%
\begin{equation*}
\left( y^{1},y^{2},y^{3},y^{4}\right) =(
V^{(1)},V^{(2)},V^{(3)},V^{(4)}) :=V,\quad V^{(i)}=\frac{dx^{(i)}}{dt},
\end{equation*}%
\begin{equation*}
v^{2}=|V|^{2}=-( V^{(1)}) ^{2}+( V^{(2)}) ^{2}+(
V^{(3)}) ^{2}+( V^{(4)}) ^{2},
\end{equation*}%
lead to the following  equations%
\begin{equation}
\dfrac{dV^{(i)}}{dt}+\frac{4\gamma v^{2}\left( 1+3\gamma ^{2}v^{2}\right) }{%
\left( 1+2\gamma ^{2}v^{2}\right) \left( 1+6\gamma ^{2}v^{2}\right) }\left(
\gamma _{s}V^{(s)}V^{(i)}\right) -\frac{\gamma v^{4}}{1+2\gamma ^{2}v^{2}}%
\gamma _{i}\eta _{ii}=0,  \label{eq-of-motin}
\end{equation}%
where $t$ is some evolution parameter.
In detailed form the equations (\ref{eq-of-motin}) read as%
\begin{equation}
\left\{
\begin{array}{c}
\dfrac{dV^{(1)}}{dt}+\frac{4\mathbf{\gamma }v^{2}\left( 1+3\mathbf{\gamma }%
^{2}v^{2}\right) }{\left( 1+2\mathbf{\gamma }^{2}v^{2}\right) \left( 1+6%
\mathbf{\gamma }^{2}v^{2}\right) }(\gamma _{s}V^{(s)})V^{(1)}+\frac{\mathbf{%
\gamma }v^{4}}{1+2\mathbf{\gamma }^{2}v^{2}}\gamma _{1}=0\medskip \\
\dfrac{dV^{(2)}}{dt}+\frac{4\mathbf{\gamma }v^{2}\left( 1+3\mathbf{\gamma }%
^{2}v^{2}\right) }{\left( 1+2\mathbf{\gamma }^{2}v^{2}\right) \left( 1+6%
\mathbf{\gamma }^{2}v^{2}\right) }(\gamma _{s}V^{(s)})V^{(2)}-\frac{\mathbf{%
\gamma }v^{4}}{1+2\mathbf{\gamma }^{2}v^{2}}\gamma _{2}=0\medskip \\
\dfrac{dV^{(3)}}{dt}+\frac{4\mathbf{\gamma }v^{2}\left( 1+3\mathbf{\gamma }%
^{2}v^{2}\right) }{\left( 1+2\mathbf{\gamma }^{2}v^{2}\right) \left( 1+6%
\mathbf{\gamma }^{2}v^{2}\right) }(\gamma _{s}V^{(s)})V^{(3)}-\frac{\mathbf{%
\gamma }v^{4}}{1+2\mathbf{\gamma }^{2}v^{2}}\gamma _{3}=0\medskip \\
\dfrac{dV^{(4)}}{dt}+\frac{4\mathbf{\gamma }v^{2}\left( 1+3\mathbf{\gamma }%
^{2}v^{2}\right) }{\left( 1+2\mathbf{\gamma }^{2}v^{2}\right) \left( 1+6%
\mathbf{\gamma }^{2}v^{2}\right) }(\gamma _{s}V^{(s)})V^{(4)}-\frac{\mathbf{%
\gamma }v^{4}}{1+2\mathbf{\gamma }^{2}v^{2}}\gamma _{4}=0.%
\end{array}%
\right.  \label{eq-of-motion-detailed}
\end{equation}

Note that, in the simplest case of the uniform medium with a constant refractive index $n(x)=n_{0}$, the above  equations become
\begin{eqnarray*}
\dfrac{dV^{(i)}}{dt} &=&0\Leftrightarrow V=(V^{(1)},V^{(2)},V^{(3)},V^{(4)})=%
\text{constant}\Leftrightarrow \\
\frac{dx^{(i)}}{dt} &=&V^{(i)}\Leftrightarrow
x(t)=(V^{(1)}t+x_{0}^{(1)},V^{(2)}t+x_{0}^{(2)},V^{(3)}t+x_{0}^{(3)},V^{(4)}t+x_{0}^{(4)});
\end{eqnarray*}%
in this case, the geodesics are the straight lines.

\section{Nonuniform nondispersive medium}

We will consider a nonuniform dispersionless medium, whose refractive
index $n(x)$ depends only on space coordinates $(x^{(2)},x^{(3)},x^{(4)})=(X,Y,Z)$.

\begin{figure}[hbt]
\begin{center}
(a) \hspace*{3.cm} (b) \hspace*{3.cm} (c) \\[0pt]
\includegraphics[width=3.7 cm,height=2.5 cm,angle=0] {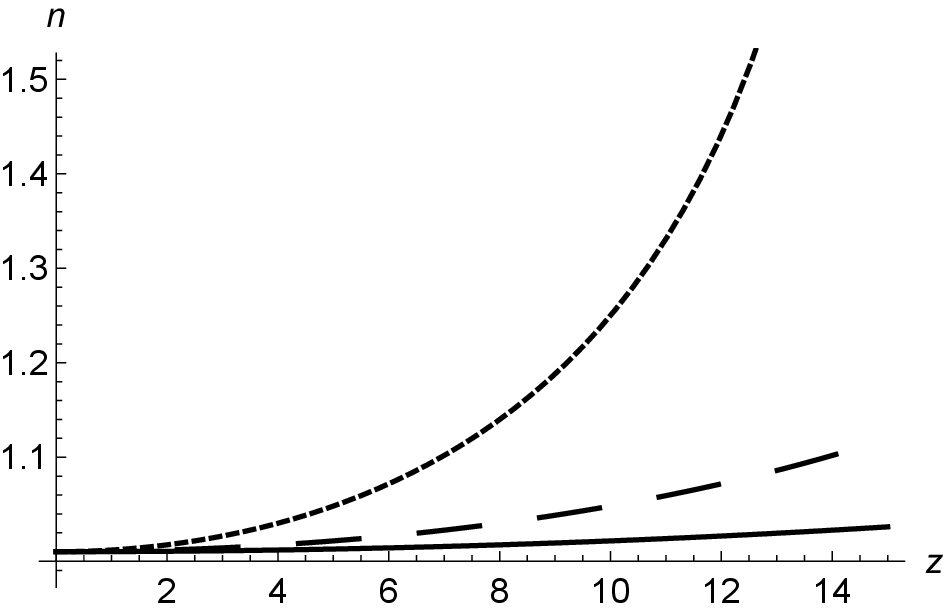} \hspace{0.1cm} 
\includegraphics[width=3.7 cm,height=2.5 cm,angle=0]{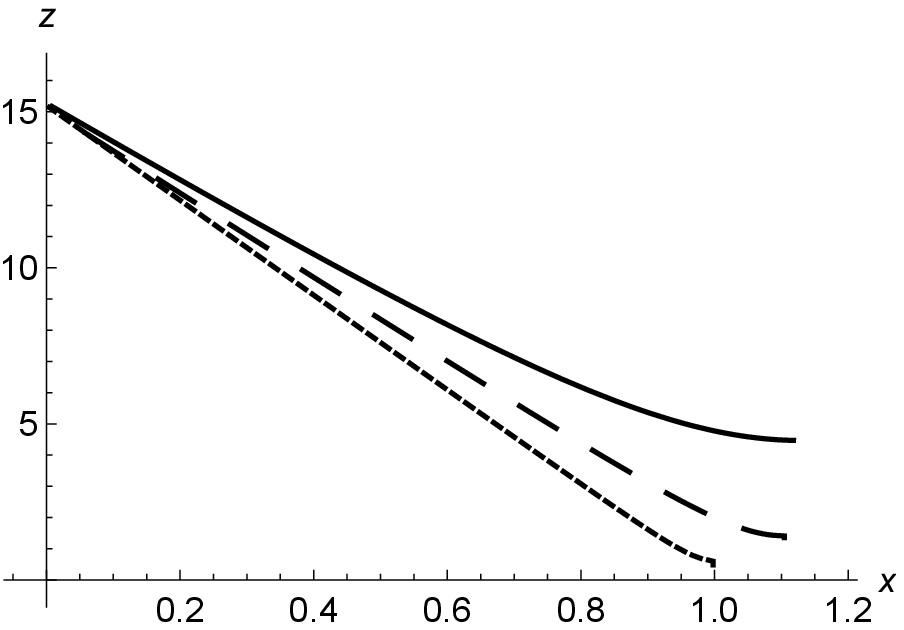} \hspace{0.1cm}
\includegraphics[width=3.7 cm,height=2.5 cm,angle=0]{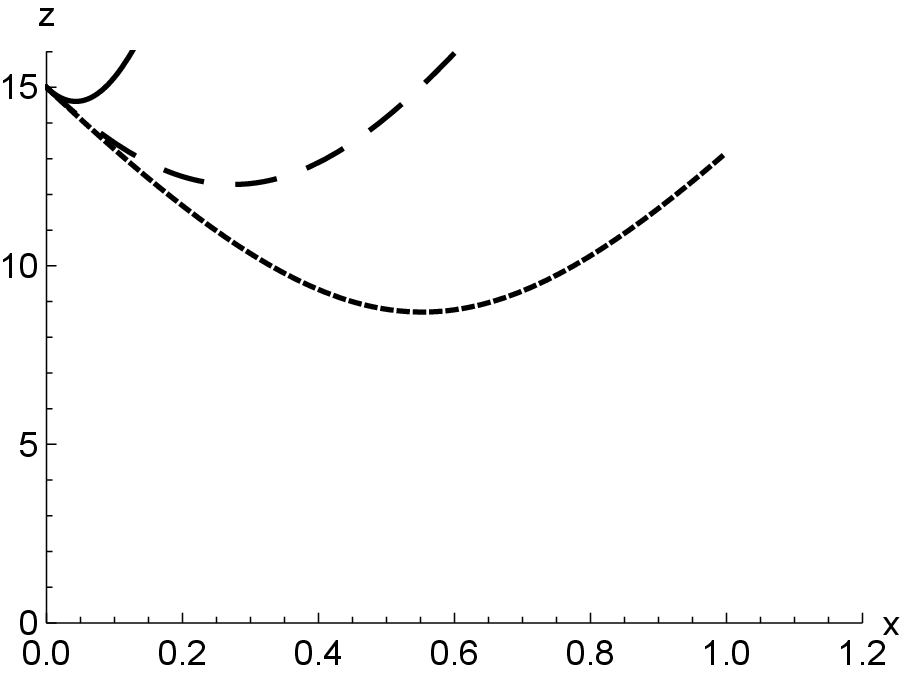}
\end{center}
\caption{(a) Dependence of the refractive index $n$ on coordinate
$Z$ at different values of $\protect\gamma_4$; (b) and (c) Trajectory of a ray. We use the following parameters: $c=1$;
$\protect\gamma_4$=0.06 (dotted line), 0.03 (dashed line), 0.01
(solid line). In (b) $c_1=-0.1$; $c_2=1$  (dotted line), 0.2
(dashed line), 0.02 (solid line). In (c) $c_1=0$;
$Z_0=15$; $X_0=0$; $V^{(4)}_0=-0.9$; $V^{(2)}_0=0.05$}
\label{fig1}
\end{figure}

Let $\gamma =\gamma _{4}x^{(4)}=\gamma _{4}Z$, this means that the function $%
\gamma (x)$ linearly increases on the $Z$-axis\footnote{%
By the symmetry of the system (\ref{eq-of-motion-detailed}), we
can treat by analogy the following similar cases: $\gamma =\gamma
_{3}Y$ or $\gamma =\gamma _{2}X.$}. From the relation
(\ref{gamman}), one can find the explicit dependence of the
refractive index on Z:
$$n^2=\frac{1}{1-c^2\gamma_4^2 Z^2}.$$
The shapes of this dependence at different $\gamma_4$ are shown in
Figure \ref{fig1}a.

Then the system (\ref{eq-of-motion-detailed}) takes the form
\begin{equation}
\begin{split}
\frac{dV^{(1)}}{dt}+\frac{4\gamma v^{2}(1+3\gamma ^{2}v^{2})}{(1+2\gamma
^{2}v^{2})(1+6\gamma ^{2}v^{2})}\gamma _{4}V^{(1)}V^{(4)}& =0, \\
\frac{dV^{(2)}}{dt}+\frac{4\gamma v^{2}(1+3\gamma ^{2}v^{2})}{(1+2\gamma
^{2}v^{2})(1+6\gamma ^{2}v^{2})}\gamma _{4}V^{(2)}V^{(4)}& =0, \\
\frac{dV^{(3)}}{dt}+\frac{4\gamma v^{2}(1+3\gamma ^{2}v^{2})}{(1+2\gamma
^{2}v^{2})(1+6\gamma ^{2}v^{2})}\gamma _{4}V^{(3)}V^{(4)}& =0, \\
\frac{dV^{(4)}}{dt}+\frac{4\gamma v^{2}(1+3\gamma ^{2}v^{2})}{(1+2\gamma
^{2}v^{2})(1+6\gamma ^{2}v^{2})}\gamma _{4}(V^{(4)})^{2}-\frac{\gamma v^{4}}{%
1+2\gamma ^{2}v^{2}}\gamma _{4}& =0.
\end{split}
\label{sys0}
\end{equation}%
For shortness, in the following we use the notation $v^{2}\equiv
W$. From (\ref{sys0}) we can derive the following equations for
variables $W$ and $V^{(4)}$:
\begin{equation}
\begin{split}
\frac{dW}{dt}+\frac{8\gamma (1+3\gamma ^{2}W)}{(1+2\gamma ^{2}W)(1+6\gamma
^{2}W)}\gamma _{4}V^{(4)}W^{2}-\frac{2\gamma }{1+2\gamma ^{2}W}\gamma
_{4}V^{(4)}W^{2}& =0, \\
\frac{dV^{(4)}}{dt}+\frac{4\gamma (1+3\gamma ^{2}W)}{(1+2\gamma
^{2}W)(1+6\gamma ^{2}W)}\gamma _{4}(V^{(4)})^{2}W-\frac{\gamma }{1+2\gamma
^{2}W}\gamma _{4}W^{2}& =0,
\end{split}
\label{sys2}
\end{equation}%
or differently
\begin{equation}
\begin{split}
\frac{dW}{dt}+\frac{6\gamma \gamma _{4}}{1+6\gamma ^{2}W}V^{(4)}W^{2}&
=0,\qquad \qquad \qquad  \\
\frac{dV^{(4)}}{dt}+\frac{4\gamma (1+3\gamma ^{2}W)}{(1+2\gamma
^{2}W)(1+6\gamma ^{2}W)}\gamma _{4}(V^{(4)})^{2}W-\frac{\gamma }{1+2\gamma ^{2}W}%
\gamma _{4}W^{2}& =0.
\end{split}
\label{sys3}
\end{equation}
Taking into account the identities
\begin{equation*}
\gamma =\gamma _{4}x^{(4)},\quad V^{(4)}=\frac{dx^{(4)}}{dt},\quad
x^{(4)}= Z,
\end{equation*}%
we rewrite the equations (\ref{sys3}) as follows
\begin{equation}
\begin{split}
\frac{dW}{dt}+\frac{6Z\gamma _{4}^{2}}{1+6\gamma _{4}^{2}Z^{2}W}W^{2}\frac{dZ%
}{dt}& =0,\qquad \qquad \qquad \qquad \quad  \\
\frac{d^{2}Z}{dt^{2}}+\frac{4\gamma _{4}Z(1+3\gamma _{4}^{2}Z^{2}W)}{%
(1+2\gamma _{4}^{2}Z^{2}W)(1+6\gamma _{4}^{2}Z^{2}W)}\gamma _{4}\left( \frac{%
dZ}{dt}\right) ^{2}W-\frac{\gamma _{4}Z}{1+2\gamma _{4}^{2}Z^{2}W}\gamma
_{4}W^{2}& =0.
\end{split}
\label{sys4}
\end{equation}

To resolve the first equation in (\ref{sys4}), let us make two
substitutions:
\begin{equation*}
3\gamma _{4}^{2}Z^{2}=F_{1},\quad \frac{1}{W}=F_{2}\quad \Longrightarrow
\quad F_{1}^{\prime }=6\gamma _{4}^{2}ZZ^{\prime },\quad W^{\prime }=-\frac{1%
}{F_{2}^{2}}F_{2}^{\prime },
\end{equation*}%
where the derivative over $t$ is denoted by a prime. The first equation in (%
\ref{sys4}) takes the form
\begin{equation*}
-F_{2}^{\prime }+F_{1}^{\prime }\frac{1}{1+2F_{1}/F_{2}}=0.
\end{equation*}%
Making an additional substitution $F_{1}/F_{2}=G$, we arrive at a
differential equation with separable variables:
\begin{equation*}
\frac{F_{2}^{\prime }}{F_{2}}=\frac{G^{\prime }}{1+G}\quad
\Longrightarrow \quad G=c_{1}F_{2}-1.
\end{equation*}%
Further turning to the initial variables $Z$ and $W$, we find the
following relation between the variables $Z$ and $W$:
\begin{equation}
Z^{2}=\frac{1}{3\gamma _{4}^{2}}\frac{c_{1}-W}{W^{2}},  \label{sol1}
\end{equation}%
where $c_{1}$ is an arbitrary constant. From the physical point of
view, the variable $Z$ should be real one, so the difference
$c_{1}-W$ has to be positive. At a chosen metric signature
$v^{2}=W=-(V^{(1)})^2+(V^{(2)})^2+(V^{(3)})^2+(V^{(4)})^2$ and
assuming  $V^{(1)}\equiv c^{2}$, where $c$ is the velocity of the
light in the vacuum, one should conclude that $W<0$. From the
other side, $Z^2>0$, so the difference $c_{1}-W$ should satisfy
the requirement of positiveness: $c_1>W$.

From the relation (\ref{sol1}) it follows the expression for $W$:
\begin{equation}
W=\frac{-1\mp \sqrt{12c_{1}\gamma _{4}^{2}Z^{2}+1}}{6\gamma
_{4}^{2}Z^{2}}. \label{sol2}
\end{equation}%
Let us introduce  a new variable $f$ defined as
\begin{equation}
f^{2}=12c_{1}\gamma _{4}^{2}Z^{2}+1.  \label{fZ}
\end{equation}%
Then the formula (\ref{sol2}) takes the form:
\begin{equation}
W=\frac{2c_{1}(f-1)}{f^2-1}; \label{sol2a}
\end{equation}%
here the negative values of $f$ correspond to a sign "--" in
(\ref{sol2}), while positive values of $f$ are in the range of $W$
defined by a sign "+". Taking into account $W<0$, one finds that
$f<1$. As $Z^2$ should be positive, the following restriction on
$f$ follows from the formula (\ref{fZ}): at  $c_1>0$, $f^2>1$; at
$c_1<0$, $f^2<1$.

Substituting $W$ from (\ref{sol2}) in the second equation of
(\ref{sys4}), we get the nonlinear differential equation of the
second order for the variable $Z$:
\begin{equation}
Z^{\prime \prime }+\frac{12c_{1}\gamma _{4}^{2}Z}{12c_{1}\gamma
_{4}^{2}Z^{2}+2\sqrt{12c_{1}\gamma _{4}^{2}Z^{2}+1}+1}Z^{\prime 2}-\frac{%
\left( \sqrt{12c_{1}\gamma _{4}^{2}Z^{2}+1}-1\right)
^{2}}{12\gamma _{4}^{2}Z^{3}\left( \sqrt{12c_{1}\gamma
_{4}^{2}Z^{2}+1}+2\right) }=0; \label{eqZ}
\end{equation}%
in terms of new variable $f$ we have
\begin{equation}
f^{\prime \prime }f+\frac{\left( f^{3}-2f-2\right) }{(f+2)\left(
f^{2}-1\right) }\left( f^{\prime }\right)
^{2}-\frac{12c_{1}^{2}\gamma _{4}^{2}(f-1)}{(f+1)(f+2)}=0.
\label{eqf}
\end{equation}%
Taking in mind the identity
\begin{equation*}
\gamma _{4}^{2}Z^{2}=\gamma ^{2}=\frac{1}{c^{2}}(1-n^{-2}),
\end{equation*}%
from (\ref{fZ}) one finds
\begin{equation*}
12(1-n^{-2})=f^{2}-1,\quad f^{2}=13-12n^{-1}.
\end{equation*}%
At $n=1$ we have $f=1$. The increasing of the refractive index $n$
leads to the rising of the variable $f$, however we should
remember that the value $f$ is restricted by the inequality
$f^{2}<13$, or $-13<f<13$. Taking in mind the previously
determined restrictions, one get $-13<f<-1$ at $c_1>0$ and
$-1<f<1$ at $c_1<0$.

Now, we will solve the equation (\ref{eqf}). Because it does not
contain the variable $t$ explicitly, we can reduce the order of
equation by means of the substitution
\begin{equation*}
f^{\prime }\rightarrow p,\quad f^{\prime \prime }\rightarrow
pp_{f}^{\prime },
\end{equation*}%
where $p_{f}^{\prime }=dp/df.$ In this way, we obtain
\begin{equation}
-\frac{12c_{1}^{2}(f-1)\gamma _{4}^{2}}{(f+1)(f+2)}+\frac{\left(
f^{3}-2f-2\right) p^{2}}{(f+2)\left( f^{2}-1\right)
}+fpp_{f}^{\prime }=0. \label{eqp}
\end{equation}%
The last equation transforms to a nonhomogeneous differential
equation by means of the substitution $p^{2}\rightarrow K$:
\begin{equation}
K_{f}^{\prime }+\frac{2\left( f^{3}-2f-2\right) }{f(f+2)\left(
f^{2}-1\right) }K-\frac{24c_{1}^{2}\gamma
_{4}^{2}(f-1)}{f(f+1)(f+2)}=0, \label{eqK}
\end{equation}%
where $K_{f}^{\prime }=dK/df.$ Whence it follows the solution
\begin{equation}
K=f^{\prime 2}=\frac{1}{f^{2}(f+2)^{2}}\left[ 24c_{1}^{2}\gamma
_{4}^{2}\left( f^{3}-1\right) -c_{2}\left( f^{2}-1\right) \right]
. \label{solK}
\end{equation}%
The last equation can be resolved implicitly in terms of the elliptic integrals $%
E[\phi |m]$ and $F[\phi |m]$ \cite{Byrd}:

\begin{equation*}
t=-\frac{\sqrt{-c_{2}f^{2}+c_{2}+\Gamma ^{2}\left( f^{3}-1\right) }%
}{3\Gamma ^{2}}
\end{equation*}%
\begin{equation*}
\times \left\{ \frac{-4c_{2}-2\Gamma
^{2}(f+5)}{f-1}-\frac{iB\left( c_{2}+3\Gamma ^{2}\right)
\sqrt{\left( \Gamma ^{2}(2f+1)-c_{2}\right)
{}^{2}-A}}{\sqrt{f-1}\left( c_{2}(-f-1)+\Gamma ^{2}\left(
f^{2}+f+1\right) \right) } \right.
\end{equation*}%
\begin{equation*}
\times \left( \frac{2\left( \sqrt{A}-3c_{2}+3\Gamma ^{2}\right)
}{\left(
2c_{2}-3\Gamma ^{2}\right) }F\left[ i\sinh ^{-1}\left( \frac{2B}{\sqrt{f-1}}%
\right) |\frac{-3\Gamma ^{2}+c_{2}+\sqrt{A}}{-3\Gamma ^{2}+c_{2}-\sqrt{A}}%
\right] \right.
\end{equation*}%
\begin{equation}
\left. \left. +4E\left[ i\sinh ^{-1}\left( \frac{2B}{\sqrt{f-1}}\right) |%
\frac{-3\Gamma ^{2}+c_{2}+\sqrt{A}}{-3\Gamma
^{2}+c_{2}-\sqrt{A}}\right] \right) \right\} ,
\end{equation}%
where $E[\phi |m]$ stands for the elliptic integral of the second kind, $%
F[\phi |m]$ is the elliptic integral of the first kind. In the
above formula the following parameters are used:
\begin{equation}
\Gamma ^{2}=24c_{1}^{2}\gamma _{4}^{2},\qquad A=\left(
c_{2}-\Gamma ^{2}\right) \left( c_{2}+3\Gamma ^{2}\right) ,\qquad
B=\sqrt{\frac{c_{2}-\frac{3\Gamma ^{2}}{2}}{\sqrt{A}+c_{2}-3\Gamma
^{2}}}.
\end{equation}%
The constants $c_{2}$ and $\Gamma (c_1)$ are defined by the
initial conditions for $(Z,Z^{\prime}).$

Now, we can find the corresponding expressions for the velocities
$V^{(1)}$, $V^{(2)}$, $V^{(3)}$ and the coordinates $x^{(1)}$ and
$X=x^{(2)}$, $Y=x^{(3)}$ (see equations (\ref{sys0})).

Because the first three equations in the system (%
\ref{sys0}) have the same form, it is sufficient to solve one of them, let it be the equation for $%
V^{(2)}=\frac{dX}{dt}$. To this end, we transform the second
equation in (\ref{sys0}) to the variable $f$.
Substituting the expressions (\ref{fZ}), (\ref{sol2}) in the second equation in (%
\ref{sys0}) and taking into account that the derivative of
$V^{(2)}$ over $t$ can be represented as
\begin{equation*}
\frac{dV^{(2)}}{dt}=\frac{dV^{(2)}}{df}\frac{df}{dt}=\frac{dV^{(2)}}{df}%
f^{\prime },
\end{equation*}%
 one get the equation
\begin{equation}
\frac{dV^{(2)}}{df}+\frac{1}{(f+2)}V^{(2)}=0,  \label{eqV2f}
\end{equation}%
whose solution  is
\begin{equation}
V^{(2)}=\frac{V_{0}^{(2)}}{f+2}.  \label{solV2f}
\end{equation}%
Finally, we can find the coordinate $X$ from the equation
\begin{equation*}
\frac{dX}{dt}=V^{(2)},
\end{equation*}%
having used the identity
\begin{equation*}
\frac{dX}{dt}=\frac{dX}{df}\frac{df}{dt}=\frac{dX}{df}f^{\prime }
\end{equation*}%
and the expression (\ref{solK}). In this way we obtain the
equation
\begin{equation*}
\frac{dX}{df}=\frac{V_{0}^{(2)}f}{\sqrt{\Gamma^2(f^3-1)-c_2(f^2-1)}}.
\label{eqX}
\end{equation*}%
The solution of this equation reads %
\begin{equation}
\begin{split}
X(f) =X_0+ V_{0}^{(2)} \left[ \frac{2 \sqrt{\Gamma ^2
\left(f^3-1\right)-c_2 \left(f^2-1\right)}}{\Gamma ^2 (f-1)}
\right. \\ \left. +\frac{i B \sqrt{f-1}}{2 \Gamma ^2 \sqrt{\Gamma
^2 \left(f^3-1\right)-c_2 \left(f^2-1\right)}}
\sqrt{\left(c_2-\Gamma ^2 (2 f+1)\right){}^2-A} \right. \\
\left. \times \left(\frac{\left(\sqrt{A}-3 c_2+3 \Gamma ^2\right)
F\left(i \sinh ^{-1}\left(\frac{2 B}{\sqrt{f-1}}\right)|1-\frac{2
\sqrt{A}}{3 \Gamma ^2-c_2+\sqrt{A}}\right)}{c_2-\frac{3 \Gamma
^2}{2}} \right. \right. \\ \left. \left.+4 E\left(i \sinh
^{-1}\left(\frac{2 B}{\sqrt{f-1}}\right)|1-\frac{2 \sqrt{A}}{3
\Gamma ^2-c_2+\sqrt{A}}\right)\right) \right].
\end{split} \label{solXf}
\end{equation}%

The trajectory $Z(X)$ has the clear physical sense that it
coincides with the trajectory of ray. Expressions (\ref{solXf})
and (\ref{fZ}) define the $XZ$-projection of the trajectory of the
ray implicitly. Its behavior depends on the parameters $\Gamma$
and $c_2$ and is illustrated in Figure \ref{fig1}b. The ray
deflects onto the direction of higher values of the refractive
index and in some point total internal reflection occurs. By a
symmetry reason, this behavior remains true for any axes-direction
which influences (by increasing) the refractive index.

It should be emphasized that the obtained solution does not
fulfilled for a particular case at $c_{1}=0$ ($\Gamma=0$).  In
this case from the formula (\ref{sol1}) we obtain
\begin{equation}
Z=\frac{1}{\sqrt{3}\gamma
_{4}}\sqrt{-\frac{1}{W}}=\frac{1}{\sqrt{3}\gamma _{4}c
\sqrt{1-V^2/c^2}}, \label{sol1a0}
\end{equation}%
where $V^{(1)}=c$, $V^2=(V^{(2)})^2+(V^{(3)})^2+(V^{(4)})^2$.

From the relation (\ref{sol1}) at $c_1=0$ it follows the
expression for $W$ in the form:
\begin{equation}
W=-\frac{1}{3\gamma _{4}^{2}Z^{2}}. \label{sol20}
\end{equation}%
At substitution $W$ from (\ref{sol20}) in the second equation of
(\ref{sys4}), the nonlinear differential equation of the second
order for the variable $Z$ reduces to the follows:
\begin{equation}
\frac{d^{2}Z}{dt^{2}}-\frac{1}{ 3\gamma^2_{4} Z^3}=0. \label{eqZ0}
\end{equation}%
The solution of this equation is
\begin{equation}
Z(t)=\sqrt{\frac{t^2}{3 \gamma_{4}^2 Z_0^2}+\left(Z_0+
V^{(4)}_{0}t\right){}^2}, \label{solZ0}
\end{equation}
where $Z_0$ and $V^{(4)}_{0}$ denote coordinate $Z$ and velocity
along the Z-axis at initial moment $t=0$. Now, we can find the
corresponding expressions for the velocities $V^{(1)}$, $V^{(2)}$,
$V^{(3)}$ and the coordinates $x^{(1)}$ and $X=x^{(2)}$,
$Y=x^{(3)}$. To do this, we substitute the expressions  (\ref{sol20}) and (\ref{solZ0}) in the equations (%
\ref{sys0}) and take into account that $\gamma =\gamma
_{4}x^{(4)}=\gamma _{4}Z$. Then one obtains
\begin{equation}
\frac{dV^{(1)}}{dt}=0, \qquad \frac{dV^{(2)}}{dt}=0, \qquad
\frac{dV^{(3)}}{dt}=0. \label{eqV20}
\end{equation}%
So, we have
$$
x^{(1)}=V^{(1)}_0 t+x_{0}^{(1)}, \qquad X=V^{(2)}_0 t+X_{0},
\qquad Y=V^{(3)}_0 t+Y_{0}.
$$

The obtained trajectories at different $\gamma_{4}$ are shown in
Fig. \ref{fig1}c. As one can see, the type of the trajectories is
similarly to the obtained in the previously case.

\section{Conclusion}
 The model which describes optical effects in
the nonuniform dispersionless moving medium has been studied. When using the optical metric restricted to the Minkowski manifold, we have established the Euler-Lagrange equation system for
geodesics. We have specified the general model to the special case
when the refractive index increases along the direction $Z$.  The
exact analytical solutions of the corresponding Euler-Lagrange
equations have been constructed.  Analysis of the solutions shows
that the light beams are bending to the axes $Z$ along which the
refractive index increase.

\noindent\textbf {Acknowledgements.} The present work was developed under the auspices of the Project BRFFR No. F20RA-007, within the cooperation framework between Romanian Academy and Belarusian Republican Foundation for Fundamental Research.
Many thanks go to Professor Y.N. Obukhov, whose useful advice helped us to improve this paper.


\noindent

{\footnotesize Mircea NEAGU}

{\footnotesize Transilvania University of Bra\c{s}ov,}

{\footnotesize Department of Mathematics-Informatics,}

{\footnotesize Blvd. Iuliu Maniu, No. 50, Bra\c{s}ov, Romania.}

{\footnotesize Email: mircea.neagu@unitbv.ro}

\medskip

\noindent

{\footnotesize Nina G. KRYLOVA}

{\footnotesize Belarusian State Agrarian Technical University,}

{\footnotesize 99 Nezavisimosti Ave., 220023, Minsk, Belarus.}

{\footnotesize Email: nina-kr@tut.by}

\medskip

\noindent

{\footnotesize Elena M. OVSIYUK}

{\footnotesize Mozyr State Pedagogical University named after I.P. Shamyakin,}

{\footnotesize Mozyr, Belarus.}

{\footnotesize Email: e.ovsiyuk@mail.ru}

\medskip

\noindent

{\footnotesize  Viktor M. RED'KOV}

{\footnotesize  National Academy of Sciences of Belarus, B.I.
Stepanov Institute of Physics,}

{\footnotesize Independence ave. 68-2, Minsk, Belarus.}

{\footnotesize Email: v.redkov@ifanbel.bas-net.by}

\end{document}